\documentclass[USenglish,twocolumn]{article}

\usepackage[utf8]{inputenc}
\usepackage[big]{dgruyter}
 
\input{epsf}
\newcommand{\be}{\begin{eqnarray}}
\newcommand{\ee}{\end{eqnarray}}
\newcommand{\hMpc}{{\ifmmode{h^{-1}{\rm Mpc}}
\else{$h^{-1}$Mpc}\fi}}
  
\begin{document}

  \articletype{Research Article{\hfill}Open Access}

  \author*[1]{V. Yankelevich }

\author[2]{S. Pilipenko}

\affil[1]{Argelander Institut f\"ur Astronomie der Universit\"at Bonn, Auf dem H\"ugel 71, D-53121 Bonn, Germany, E-mail: vyankelevich@astro.uni-bonn.de}


  \affil[2]{Astro Space Center of Lebedev Physical Institute of  Russian Academy of Sciences,Profsojuznaja st. 84/32, 117997 Moscow, Russia, E-mail:spilipenko@asc.rssi.ru}

  \title{\huge Towards accurate rescaling of a halo mass function}

  \runningtitle{Rescaling of a halo mass function}


  \begin{abstract}
{We investigate the precision within which a simulated dark matter halo mass function can be rescaled to a different set of cosmological parameters. Our tests show that the accuracy almost linearly depends on the difference of the cosmological parameters and amounts to few percent in the case of WMAP5 and PLANCK parameters. The rescaling thus allows us to obtain a mass function with better precision than the one given by the Sheth-Mo-Tormen approximation and even more modern fits currently used in the literature.}
\end{abstract}
  \keywords{dark matter simulations, cosmological simulations}

  \journalname{Open Astronomy}
\DOI{DOI}
  \startpage{1}
  \received{..}
  \revised{..}
  \accepted{..}

  \journalyear{2014}
  \journalvolume{1}

\maketitle
\section{Introduction}
The abundance of dark matter haloes is now believed to be a sensitive cosmological test (e.g. \cite{vikhlinin}). The halo mass function (hereafter MF) is used as an ingredient for the clustering models of haloes and galaxies (\cite{cooray_sheth}, \cite{viero}). Also it is the subject of the ``satellite abundance'' problem (\cite{Boylan-Kolchin}). 

The most popular, useful and developed analytical model of MF prediction is the Press-Schechter method (\cite{PS}, \cite{bond}, \cite{bower}, \cite{lacey_cole_93}). It is based on the spherical collapse of a dusty matter cloud which make this model one of the simplest. However, numerical simulations of Large Scale Structure show that the Press-Schechter model predicts somewhat wrong distributions (\cite{lacey_cole}, \cite{ST}, \cite{jenkins}, \cite{efstat}, \cite{efstat_rees}, \cite{white_efstat}).
It was shown that replacing the type of the collapse can help produce MFs closer to the simulation results (\cite{DP1}, \cite{DP2}). 
One of the most popular methods is  the Sheth-Mo-Tormen (\cite{SMT}) method, where a spherical collapse is substituted by an ellipsoidal collapse. This gives rise to more accurate results, especially at high masses. The results of simulated MFs are usually represented as analytical fits (e.g.  \cite{warren},  \cite{tinker}) which nowadays have the precision of 5--10\%. There also exist more sophisticated methods of MF calculation from the initial random density field without running a simulation, e.g. the PINOCCHIO code (\cite{monaco}).

The most reliable results on the MF are now obtained with the help of large N-body simulations. However, they are costly, especially when the goal is to investigate the dependence of the MF on cosmological parameters. On the other hand, nowadays, in the era of precise cosmology, the range of parameter change is quite small, e.g. the difference in the matter density $ \Omega_{\mathrm{m}}$ between WMAP5 and Planck cosmologies is less then  0.05.

It is therefore attractive to scale the results of existing \textit{N}-body simulations instead of doing new ones. In order to do this one needs to calculate analytically how the MF changes with the small change of cosmological parameters. This calculation can be done based on the simplest Press-Schechter method. Examples of such scaling can be found in \cite{LoVerde}, \cite{Kang} and \cite{Grossi} who used a simulated MF for Gaussian initial conditions and multiplied it by a correction factor extracted from the ratio of a non-Gaussian Press-Schechter MF to a Gaussian one. A similar approach was also used by \cite{Barkana04} and \cite{Ahn} who computed an MF in different environments by taking a high precision global MF (Sheth-Mo-Tormen in \cite{Barkana04} and a simulated one in \cite{Ahn}) and multiplying it by a bias factor computed from an excursion set approach.

Despite the fact that the rescaling of MF has been used in several papers, there is no exhaustive investigation of the accuracy of this approach in the literature to date. For example, \cite{Ahn} computed the number of haloes in the mass range $10^5-10^9 M_\odot$ as a function of large-scale density and found that using the rescaling of either \textit{N}-body MF or Sheth-Mo-Tormen MF for this purpose is much more accurate than using a linear bias model with Press-Schechter MF.

The rescaling of MF may turn out to be useful for the exploration of a variety of non-standard cosmologies as well as fine-tuning the cosmological parameters to fit the observed Universe. This motivated us to investigate the precision of the rescaled mass function in this paper. For this purpose, we run a set of simulations with varying cosmological parameters: matter density $ \Omega_{\mathrm{m}}$, amplitude of the power spectrum $\sigma_8$ and slope of the initial power spectrum $n_s$. The accuracy of the rescaled MF is then expressed as a function of the difference of these parameters between the source and target cosmologies.

There is another way of rescaling: treating the MF as a function of $\sigma(M)$, the RMS of the matter density fluctuations scale with M. Press-Schechter theory implies that the dependency of the MF on $\sigma(M)$ does not depend itself on the cosmological parameters, i.e. it is universal. During the past decade this universality has been intensively checked (\cite{jenkins}, \cite{warren}, \cite{tinker}, \cite{bhattacharya}). The conclusion is that the MF may differ from the universal one by 5--10\% in case of Friends-of-Friends haloes, or 20--50\% for spherical haloes (\cite{tinker}).

Another technique of rescaling was demonstrated by \cite{angulo_white}. They created the algorithm that allows one to scale the output of a cosmological \textit{N}-body simulation carried out for one specific set of cosmological parameters so that it faithfully represents the growth of structure in a different cosmology.

The paper is organized in the following way. Firstly, we briefly outline the theoretical background on which the rescaling method is based. In Section 2, we present results and test the accuracy of the method for different cosmological models. In the last section, we present a summary of our work and discuss some astrophysical implications.

\section{Methods and Results}

\subsection{Method of MF rescaling}
Firstly, let us describe the basics of the method with a simple example. In Figure 1, we present two mass functions produced in simulations and two mass functions calculated using the Press-Schechter method for two different sets of parameters: in one case $ \Omega_{\mathrm{m}}=0.2$ ($\Omega_\Lambda=0.8$; hereafter we will omit $\Omega_\Lambda$ implying flat cosmology) and in another one $ \Omega_{\mathrm{m}}=0.3$.

\begin{figure}[tbp]
\centering
\epsfxsize=8cm \epsfbox{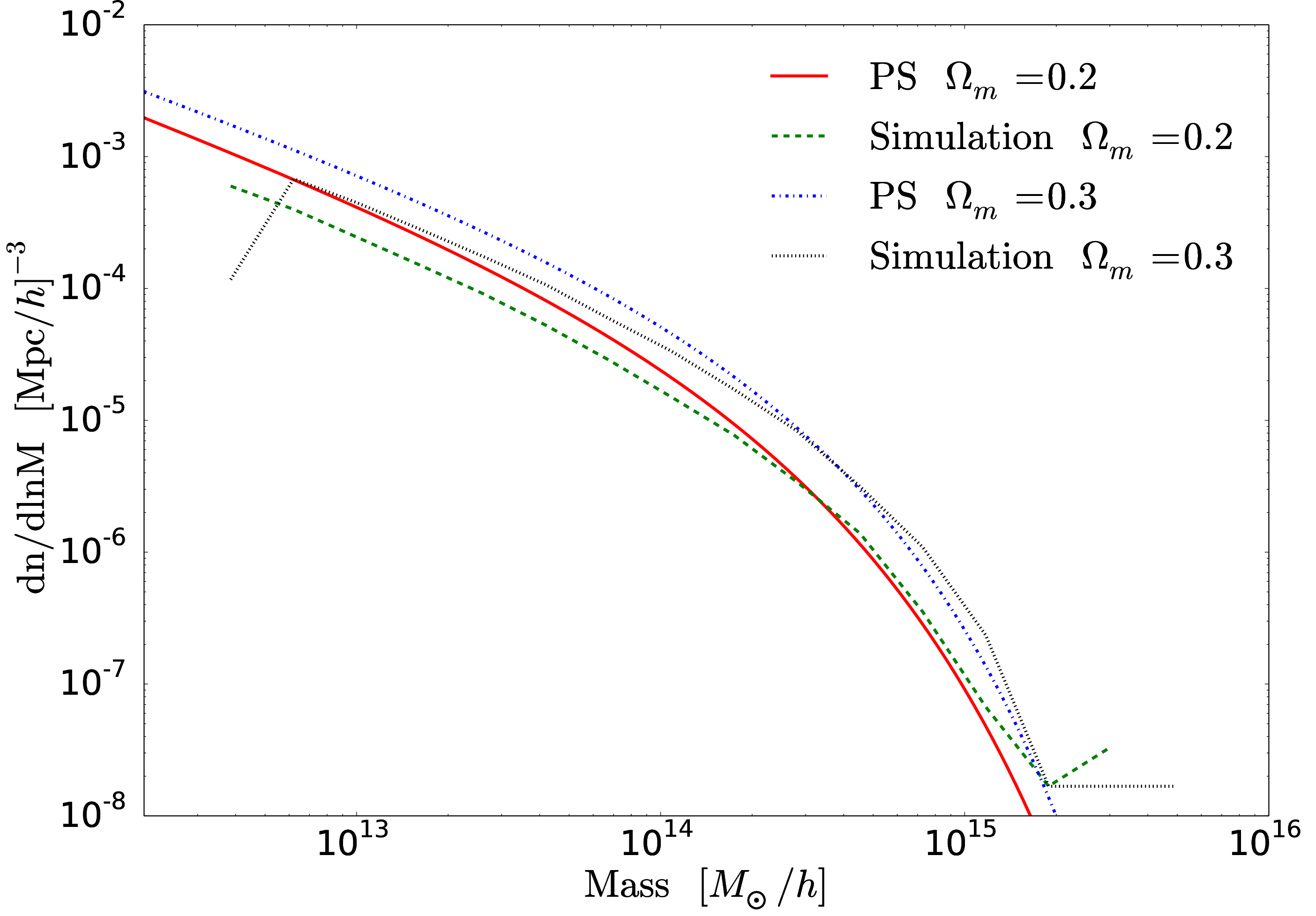}
\caption{Halo mass functions for two sets of cosmological parameters: $ \Omega_{\mathrm{m}}=0.2$ and $ \Omega_{\mathrm{m}}=0.3$ obtained with Press-Schechter method and simulations.}
\end{figure}

The values $ \Omega_{\mathrm{m}}=0.2$ and $ \Omega_{\mathrm{m}}=0.3$ were chosen for demonstration purposes. It is seen that the curve shapes are similar for theory and simulations (points near the right edge contain large errors because of the small number of haloes in this region). The similarity is demonstrated even more clearly in the Figure 2, where the ratio of two Press-Schechter MFs and two simulation MFs are shown.

\begin{figure}[tbp]
\centering
\epsfxsize=8cm \epsfbox{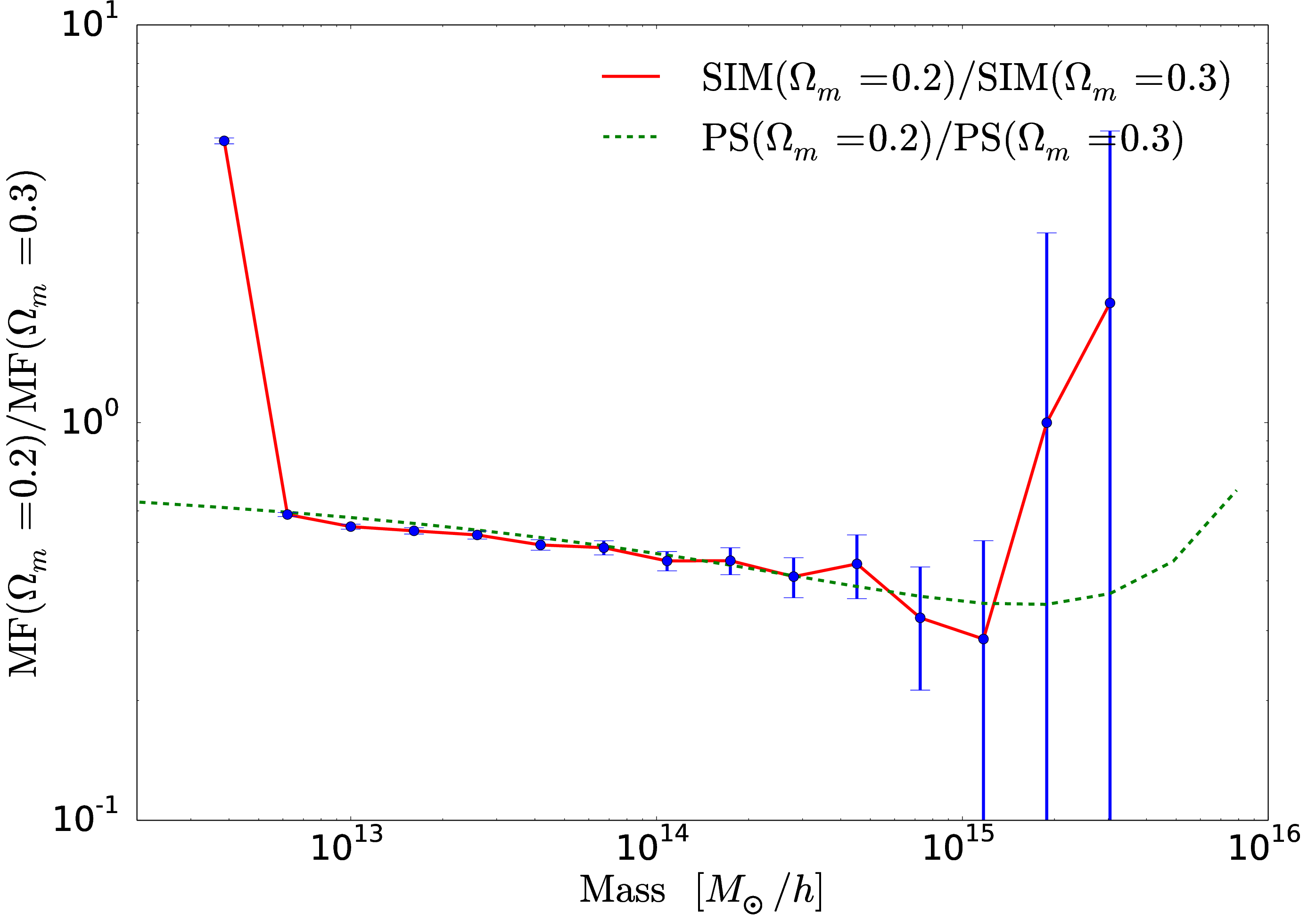}
\caption{The ratio of two Press-Schechter mass functions and two simulated mass functions.}
\end{figure}

Due to this finding, we suppose that the ratio between pairs of simulated and theoretical MFs at any given mass is equivalent and corresponds to the next equations (the dependency of MFs on mass is omitted):

The ratio between Press-Schechter MFs:
\be
\label{eq2}
\rm{R_{PS}}=\frac{\rm{PS}( \Omega_{\mathrm{m}}=0.2)}{\rm{PS}( \Omega_{\mathrm{m}}=0.3)}.
 \ee
\noindent The ratio between simulated MFs:
\be
\label{eq3}
\rm{R_{SIM}}=\frac{\rm{SIM}( \Omega_{\mathrm{m}}=0.2)}{\rm{SIM}( \Omega_{\mathrm{m}}=0.3)}.
 \ee
\noindent From the similarity of the ratios 
\be
\label{eq4}
\rm{R_{PS}}\approx \rm{R_{SIM}},
 \ee
\noindent it follows that the new MF can be expressed as
\begin{multline}
\label{eq5}
\rm{Rescaled}( \Omega_{\mathrm{m}}=0.3)=\\  \frac{\rm{PS}( \Omega_{\mathrm{m}}=0.3)}{\rm{PS}( \Omega_{\mathrm{m}}=0.2)} \cdot \rm{SIM}( \Omega_{\mathrm{m}}=0.2)
 \end{multline}


In Figure 3, we present an original simulated mass function for $ \Omega_{\mathrm{m}}=0.3$ and an extrapolation of the mass function for this set of parameters from $ \Omega_{\mathrm{m}}=0.2$ using (\ref{eq5}). They are extremely close. The error is very small mostly everywhere except for high masses. The main goal of our research was to quantify this error as a function of the difference in cosmological parameters.

\begin{figure}[tbp]
\centering
\epsfxsize=8cm \epsfbox{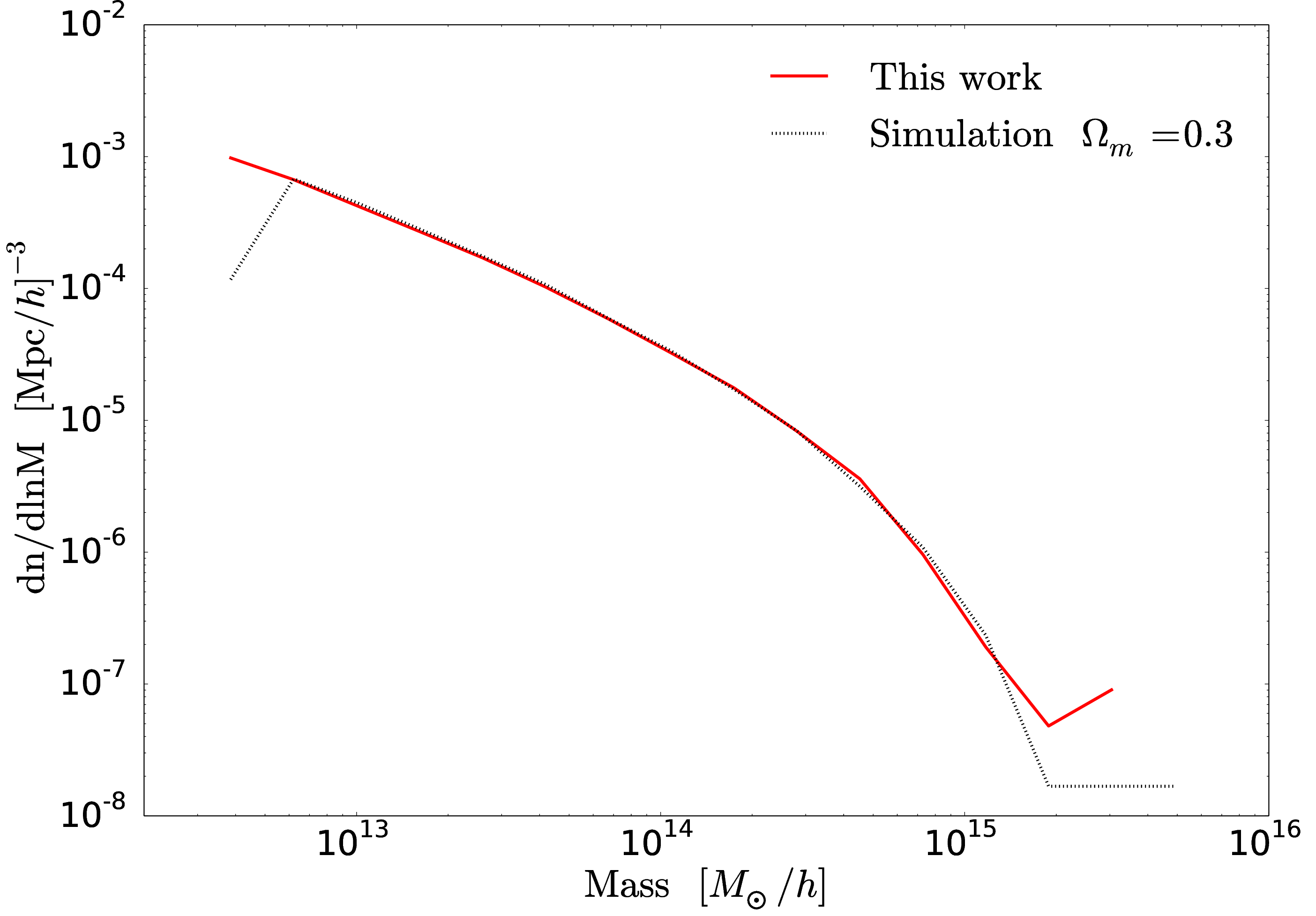}
\caption{The MF from the original simulation for $ \Omega_{\mathrm{m}}=0.3$ and the reconstructed MF by using (\ref{eq5}) for $ \Omega_{\mathrm{m}}=0.3$.}
\end{figure}

\subsection{Simulations}
To test the accuracy of rescaling, we run a set of \textit{N}-body simulations using GADGET-2 code (\cite{springel})  with dark matter only. The volume of all the simulations is a cubic box of $500 h^{-1}$ Mpc in size, the total number of particles is $512^3$ and the gravitational softening parameter is $25 h^{-1}$ kpc. For the initial conditions, we used the \textit{N}-GenIC code written by V. Springel. We held the phases of the initial velocity field the same for all of the simulations. The cosmological model is a flat Universe with $\Lambda$CDM cosmology where $ \Omega_{\mathrm{m}}+\Omega_\Lambda=1$. The simulations were run on the supercomputer of the Nuclear Physics and Astrophysics Division of P.N. Lebedev Physical Institute.

For finding the haloes, we used the Amiga's Halo Finder (AHF) (\cite{gill}, \cite{knebe}). The minimum number of particles in a halo is set to $60$ and the total number of haloes in each simulation is about $10^5$ at $z=0$. For the MF construction, the Friends-of-Friends (\cite{FOF}) method is used quite often, e.g. in \cite{warren}, so we also used the RockStar halo finder (\cite{Behroozi}). The virial overdensity $\Delta$ is calculated for each set of cosmological parameters using a spherical top-hat collapse model (\cite{gross}). Only the isolated haloes are used.

We vary three cosmological parameters, $ \Omega_{\mathrm{m}}$, $\sigma_8$ and $n_s$. Our set of simulations is summarized in Table 1. In addition to our own simulations, we use two publicly available simulations: Bolshoi and BolshoiP (\cite{klypin}). The halo catalogues were obtained from the COSMOSIM database (www.cosmosim.org). The haloes were identified using the standard overdensity criterion with $\Delta = 360\cdot \rho_{back}$ with the spherical overdensity-based BDM halo finder.

\subsection{The testing of the method accuracy}

\begin{table}[tbp]
\centering
\label{ref:tab1}
\begin{tabular}{|l|llll|}
\hline
Name & $ \Omega_{\mathrm{m}}$ & $\Omega_\Lambda$ & $\sigma_8$ & $n_s$ \\
\hline
Sim($ \Omega_{\mathrm{m}}$=0.2) & 0.2 & 0.8 & 0.8 & 1.0 \\
Sim($ \Omega_{\mathrm{m}}$=0.25) & 0.25 & 0.75 & 0.8 & 1.0 \\
Sim($ \Omega_{\mathrm{m}}$=0.3) & 0.3 & 0.7 & 0.8 & 1.0 \\
Sim($ \Omega_{\mathrm{m}}$=0.35) & 0.35 & 0.65 & 0.8 & 1.0 \\
Sim($ \Omega_{\mathrm{m}}$=0.4) & 0.4 & 0.6 & 0.8 & 1.0 \\
Sim($\sigma_8$=0.9) & 0.25 & 0.75 & 0.9 & 1.0 \\
Sim($\sigma_8$=1.0) & 0.25 & 0.75 & 1.0 & 1.0 \\
Sim($n_s$=0.9) & 0.25 & 0.75 & 0.8 & 0.9 \\
Sim($n_s$=0.8) & 0.25 & 0.75 & 0.8 & 0.8 \\
Bolshoi & 0.27 & 0.73 & 0.82 & 0.95 \\
BolshoiP & 0.30711 & 0.69289 & 0.82 & 0.96 \\
\hline
\end{tabular}
\caption{Parameters of simulations used in tests}
\end{table}

In order to measure the accuracy of the method, we use the maximum difference between two cumulative MFs. In simulations and the Press-Schechter method, we obtain differential MFs (here the division into $1000$ bins is used), integrate them and then interpolate for the same mass interval. Finally, we take a maximum of the difference between the original and calculated MFs divided by the original MF. The maximal value defined this way we call the error of the method and it is shown in Figures 4-6. In each plot, the errors for the rescaled MF are presented for pure Press-Schechter, Sheth-Mo-Tormen, Tinker (2008), Angulo (2012), Crocce (2010), Courtin (2011), Bhattacharya (2011) and Del Popolo (2017)  fits. 

From the Figures 4-6 we see that for the small change in cosmological parameters the method of rescaling has much better accuracy than analytical MF fits. The accuracy of the rescaled MF also can be quantified in the following way:

\begin{equation}
\label{eqOmega}
\Delta MF=36\%\; \Delta  \Omega_{\mathrm{m}},
\end{equation}

\begin{figure}[tbp]
\centering
\epsfxsize=8cm \epsfbox{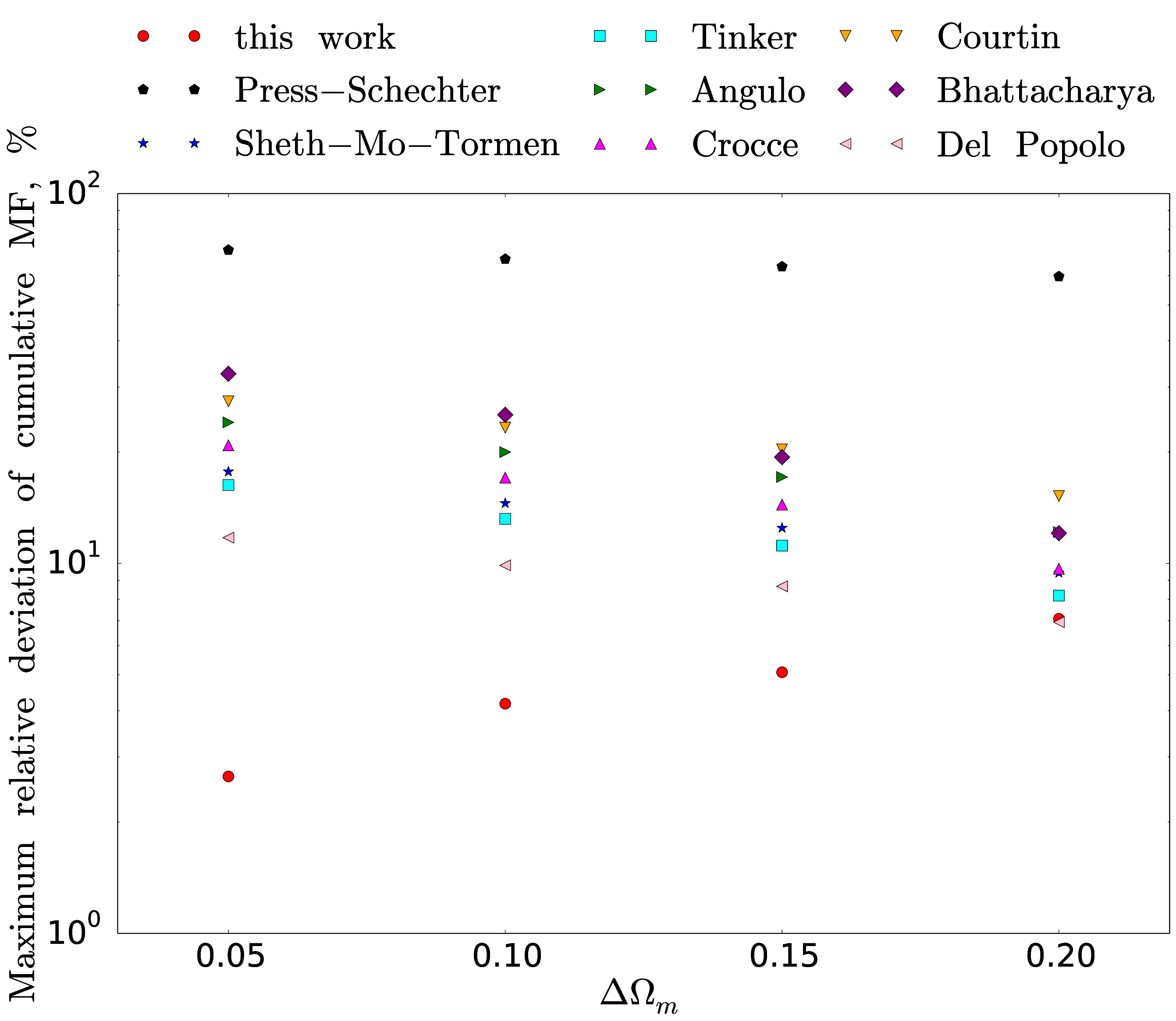}
\caption{The accuracy of MF calculation methods in dependence upon the $ \Omega_{\mathrm{m}}$ difference.}
\end{figure}

\begin{equation}
\Delta MF=18\%\; \Delta n_s,
\end{equation}

\begin{figure}[tbp]
\centering
\epsfxsize=8cm \epsfbox{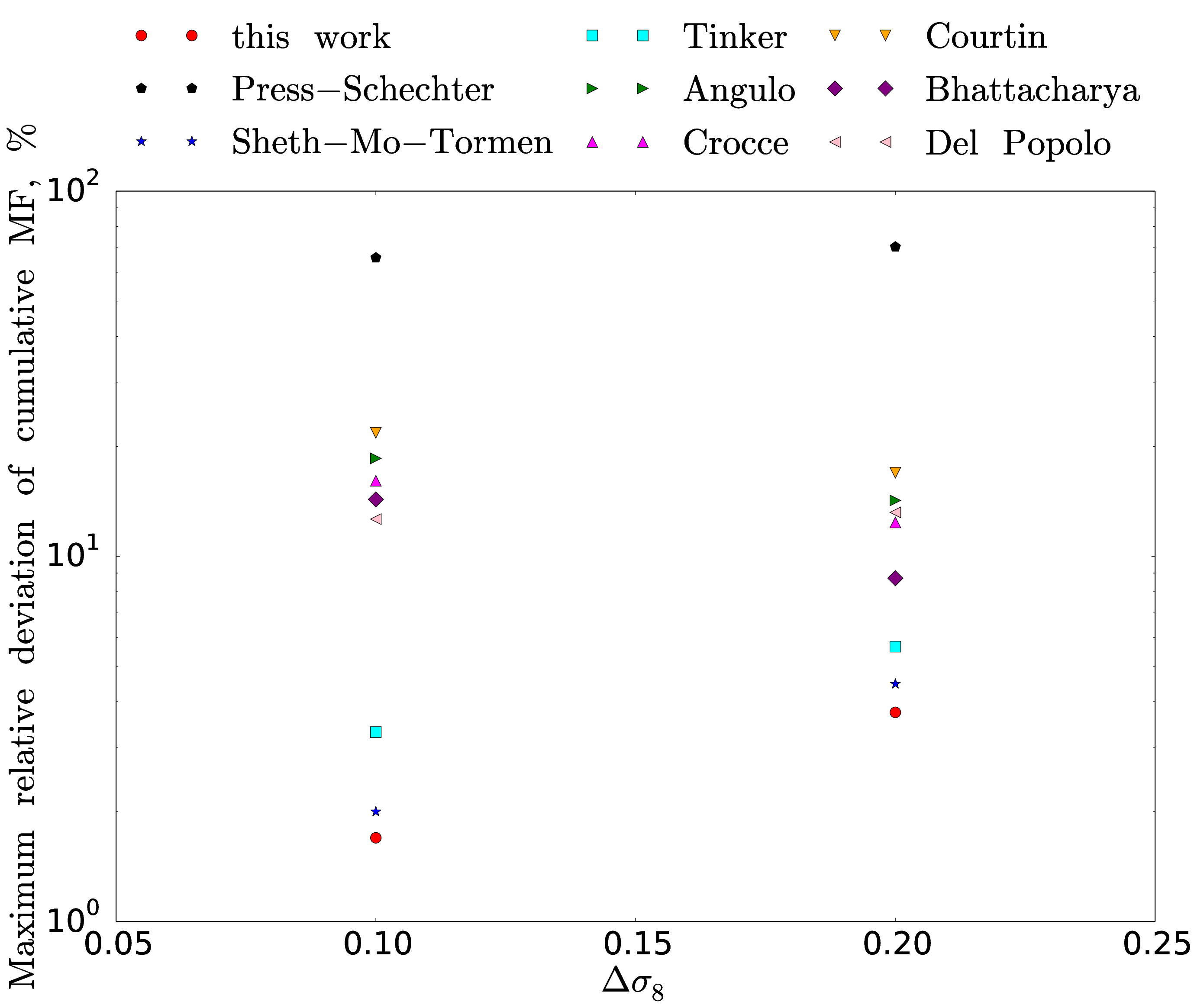}
\caption{The accuracy of MF calculation methods in dependence upon the $\sigma_8$ difference.}
\end{figure}

\begin{equation}
\label{eqSigma8}
\Delta MF=30\%\; \Delta \sigma_8.
\end{equation}

\begin{figure}[tbp]
\centering
\epsfxsize=8cm \epsfbox{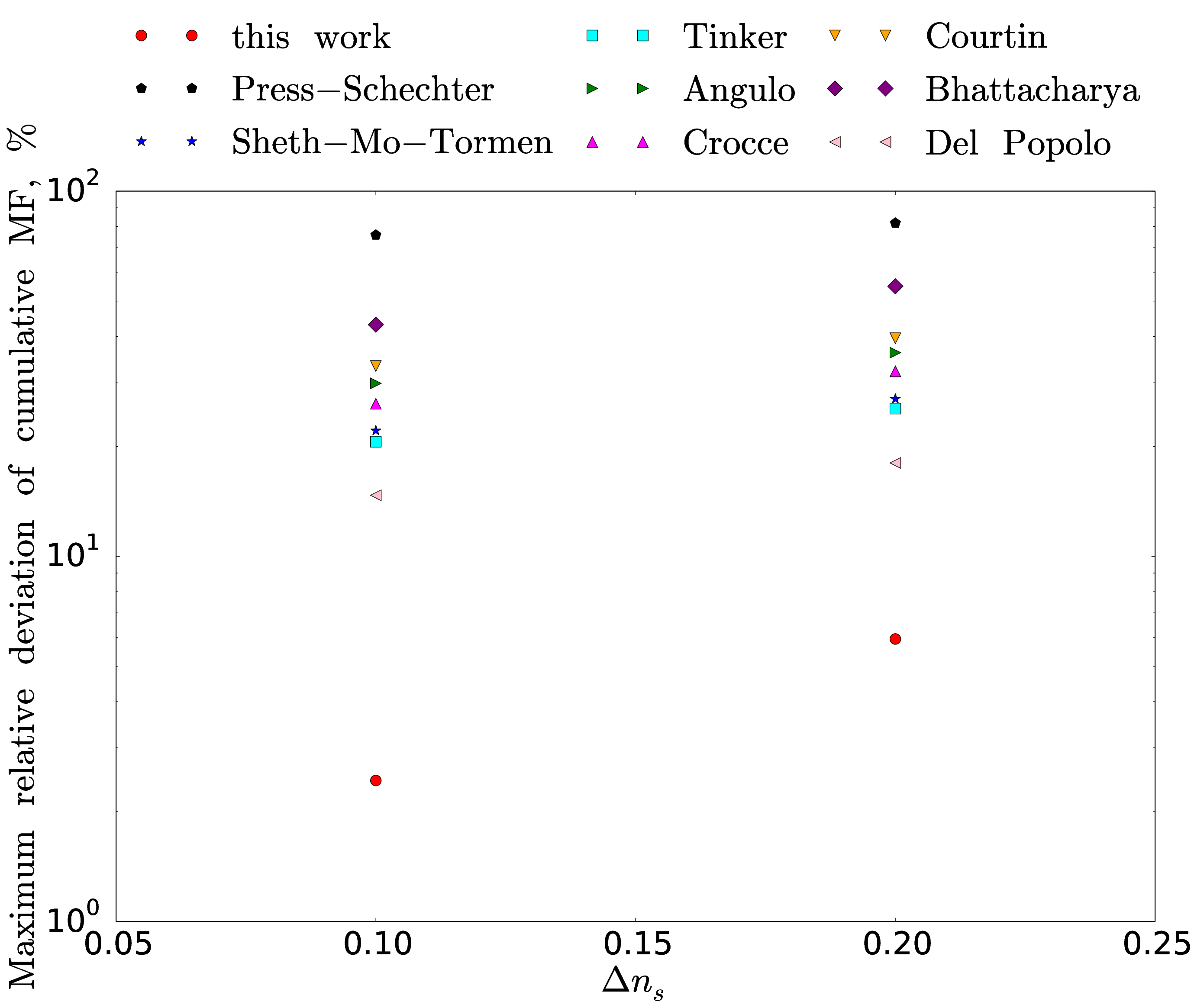}
\caption{The accuracy of MF calculation methods in dependence upon the $n_s$ difference.}
\end{figure}

\begin{figure}[tbp]
\centering
\epsfxsize=\linewidth
\epsfxsize=8cm \epsfbox{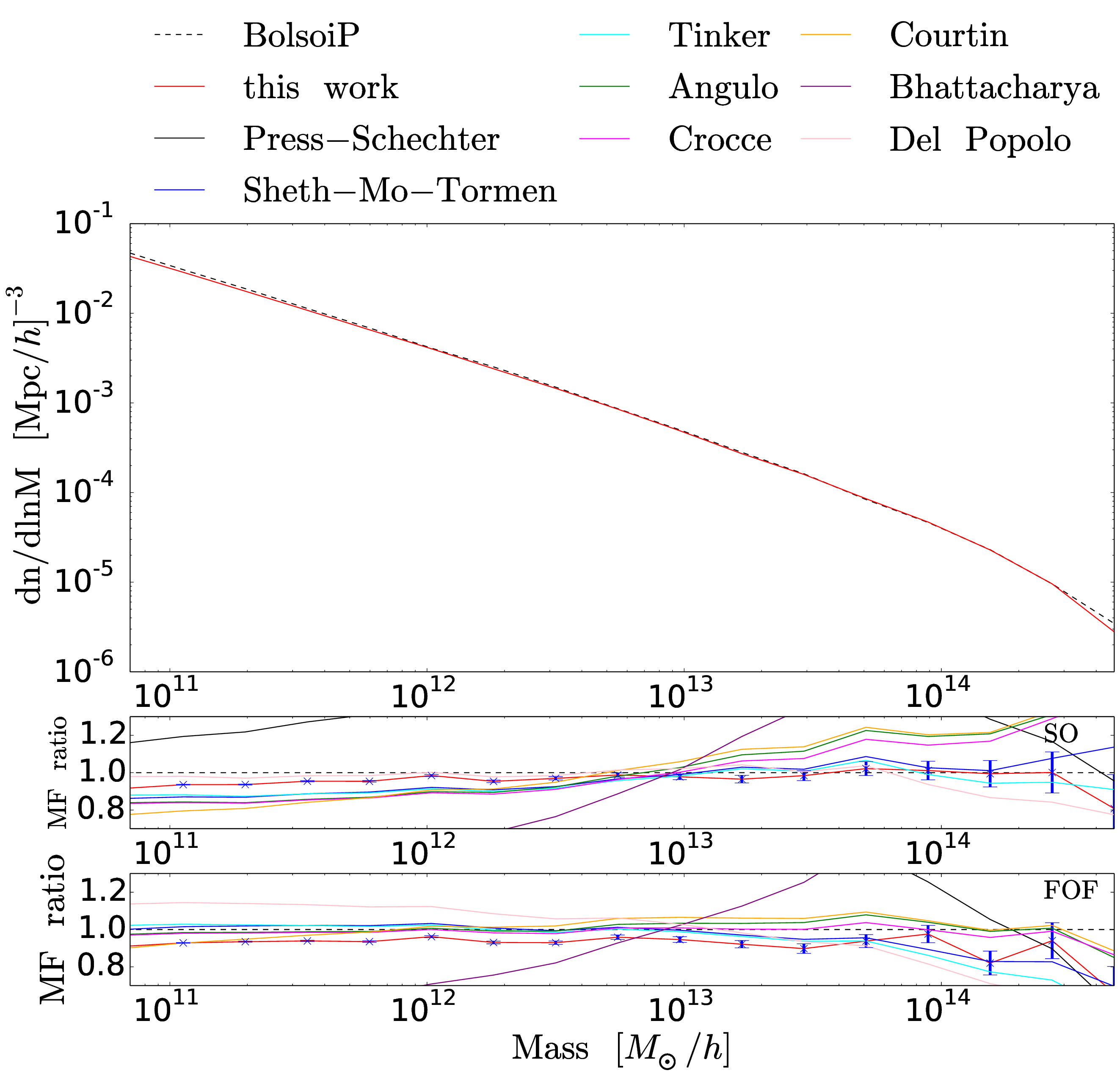}
\caption{Top: The original MF for the BolshoiP simulation and the MF calculated by rescaling using equation (\ref{eq5}) and the MF for the Bolshoi simulation for the SO case. 
\newline
Middle: The ratio of different MF approximations to the MF measured from BolshoiP simulation for the SO case.
\newline
Bottom: The ratio of different MF approximations to the MF measured from BolshoiP simulation for the FOF case.}
\end{figure}

The result of test of the rescaling method with the Bolshoi and BolshoiP simulations is shown in Figure 7. In it, the original MF for the Planck parameters together with the MF were obtained using equation (\ref{eq5}) from WMAP-based simulations are shown on the top. On the middle part the ratio of MFs is shown. The net error  defined above is 6\% for these simulations. It is 4 times higher than expected from (\ref{eqOmega}), however still much less than for pure Press-Schechter (70\%) or Sheth-Mo-Tormen (20\%). One should note that Bolshoi and BolshoiP cover two orders of magnitude larger mass range than our own simulations, so the error in the cumulative MF is being accumulated. If we consider only haloes with $M>10^{12}M_\odot$ in Bolshoi and BolshoiP, the error decreases to 2.8\%.

From Figure 7, it is seen that the rescaled MF has very good agreement compared to the simulation in the whole presented mass range. However, Del Popolo (2017), Angulo (2012), Crocce (2010) and Courtin (2011) fits can provide the same or better agreement with the simulation, but not on all masses in the range. At larger masses they overpredict (SO case) or underpredict (FOF case) the number of haloes while the rescaled MF is the most precise. According to this fact, we find the method presented in this paper more stable in all mass ranges.

We also test how our results depend on the halo finder used (FoF or spherical overdensity) (the middle and bottom parts of the Figure 7) and find that the impact of the halo identification algorithm is quite small: the error of the rescaled MF is about 20\% larger for the FoF haloes than for the spherical overdensity. This was also noticed by \cite{Mo} who found that the halo bias factor is insensitive on how the haloes are identified. Only Tinker (2008) fit is using SO, that is why the errors are bigger in the bottom part of the Figure 7 rather than on the middle, especially for higher mass. The rest of the fits are based on FoF algorithm and they show much better agreement in the bottom part of the figure.

\begin{figure}[tbp]
\centering
\epsfxsize=\linewidth
\epsfxsize=8cm \epsfbox{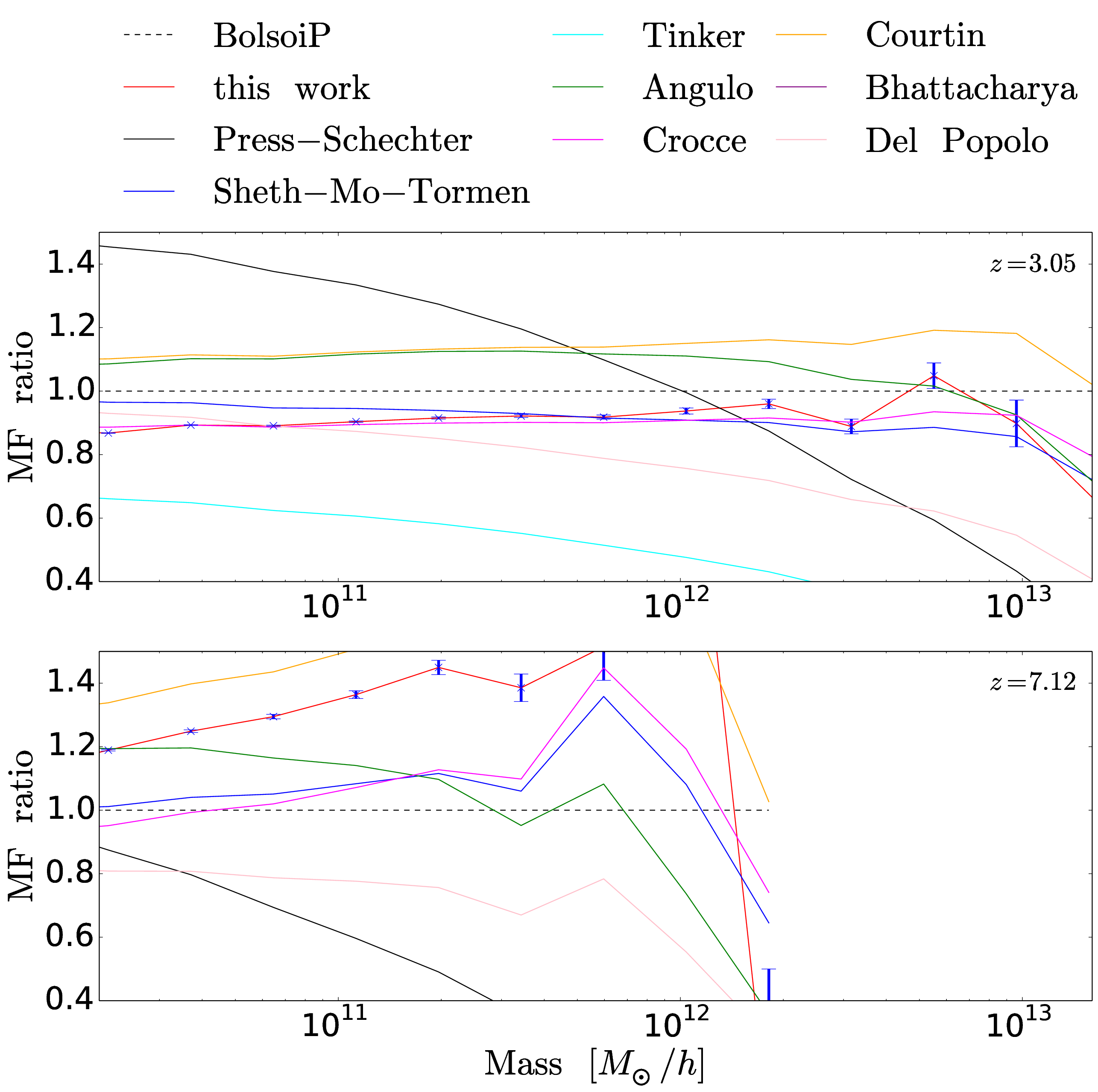}
\caption{Top: The ratio of different MF approximations to the MF measured from BolshoiP simulation for the FoF halos at $z=3.05$. 
\newline
Bottom: The ratio of different MF approximations to the MF measured from BolshoiP simulation for the FOF halos at $z=7.12$.}
\end{figure}

We test our method and theoretical fits for different redshifts, for instance, $z=3.05$ and $z=7.12$ (snapshots from BolshoiP simulation). In the Figure 8 it is clearly seen that at $z=3.05$ our method reproduces the MF almost with the same accuracy as for $z=0$ case. The same behaviour is observed for theoretical fits, except for Tinker (2008). At $z=7.12$ the method presented in this paper works somewhat worse. This can be explained by the fact that we work with high-sigma peaks at this redhift and mass range. In the Figure 8 the right boundary is fixed by data from BolshoiP simulation.

\section{Summary and Discussion}
We have investigated the accuracy of the method of calculating an MF of dark matter haloes based on the rescaling of an MF measured in simulations. The method is described by equation (\ref{eq5}). Our tests show that when the difference of cosmological parameters between the target and the source cosmologies is small, the accuracy is very high, reaching a few percent (see Figures 4--7 and equations (\ref{eqOmega})--(\ref{eqSigma8})).
This method can be applied to problems which require varying cosmological parameters, such as the fitting of observational data.

In comparison to the most accurate existing fits to simulations (e.g. \cite{tinker}, \cite{Angulo2},  \cite{Crocce}, \cite{Courtin}, \cite{bhattacharya}, \cite{dp-fit}), the rescaled MF has several advantages. Firstly, it predicts the number of haloes for the same technique of halo identification which is used in the source simulation, which may differ from one used in the available fits. Secondly, we expect that it can be used for non-standard cosmological models, e.g. with warm dark matter, non-standard dark energy, etc.

Another potential application of the rescaling may be the increase of precision on the high mass end. The idea is to run an existing simulation past $z=0$ in order to facilitate the growth of high mass haloes and thus increase statistics. Then the MF can be rescaled back to $z=0$ using equation (\ref{eq5}).
For example, in order to reduce the statistical errors two times one needs to increase the number of haloes four times. For masses $M>10^{15}M_\odot$, to do so we need to go in future by $z=-0.33$ or 6 Gyr (in PLANCK cosmology).

One can also try to use the universal MF approach for a similar procedure of rescaling, i.e. to express an MF from a source simulation as a function of $\sigma$ and then to substitute $\sigma(M)$ for the target cosmology. In this approach, the mass range of the resulting MF differs from that of the source simulation, which will not allow the use of this method to increase the precision at high mass end as described above. Another problem will arise in the case of WDM since $\sigma$ weakly depends on $M$ for masses smaller than the filtering mass. Therefore $M(\sigma)$ becomes undefined for large enough $\sigma$ and, hence, this approach may fail in some cases when the source cosmology is WDM.


The authors thank Mikhail Zavertyaev for providing computational time and tuning the NPAD LPI cluster. We also thank Kristin Riebe for helping with the Cosmosim database, Ravi Sheth and Cristiano Porciani for helpful discussion. The work of S. Pilipenko is supported by the Fundamental Research Program of Presidium of the RAS 7 and by grant of President of Russian Federation for Support of Leading Scientifics School NSh-6595.2016.2. 
The major part of the work was done during staying VY at the Souther Federal University, Russia and was supported by the Dynasty Foundation. VY was partly supported by the Deutsche Forschungsgemeinschaft through the Transregio 33 "The Dark Universe", partially supported through a research contract from the International Max Planck Research School (IMPRS) for Astronomy and Astrophysics at the Universities of Bonn and Cologne and partially supported by the Bonn-Cologne Graduate School for Physics and Astronomy

The CosmoSim database used in this paper is a service by the Leibniz-Institute for Astrophysics Potsdam (AIP).
The MultiDark database was developed in cooperation with the Spanish MultiDark Consolider Project CSD2009-00064.
The authors gratefully acknowledge the Gauss Centre for Supercomputing e.V. (www.gauss-centre.eu) and the Partnership for Advanced Supercomputing in Europe (PRACE, www.prace-ri.eu) for funding the MultiDark simulation project by providing computing time on the GCS Supercomputer SuperMUC at Leibniz Supercomputing Centre (LRZ, www.lrz.de).

    The Bolshoi simulations have been performed within the Bolshoi project of the University of California High-Performance AstroComputing Center (UC-HiPACC) and were run at the NASA Ames Research Center.

We also used HMFcalc (http://hmf.icrar.org/) by \cite{Murray} to verify our calculations of Press-Schechter, Sheth-Mo-Tormen, Tinker (2008), Angulo (2012), Crocce (2010), Courtin (2011) and Bhattacharya (2011) fits.

\def\apj{Astrophys.~J}
\def\apjl{Astrophys.~J.,~Lett}
\def\apjs{Astrophys.~J.,~Supplement}
\def\an{Astron.~Nachr}
\def\aap{Astron.~Astrophys}
\def\mnras{Mon.~Not.~R.~Astron.~Soc}
\def\pasp{Publ.~Astron.~Soc.~Pac}
\def\aaps{Astron.~and Astrophys.,~Suppl.~Ser}
\def\apss{Astrophys.~Space.~Sci}
\def\ibvs{Inf.~Bull.~Variable~Stars}
\def\japa{J.~Astrophys.~Astron}
\def\na{New~Astron}
\def\aspproc{Proc.~ASP~conf.~ser.}
\def\aspcs{ASP~Conf.~Ser}
\def\aj{Astron.~J}
\def\actaa{Acta Astron}
\def\araa{Ann.~Rev.~Astron.~Astrophys}
\def\caosp{Contrib.~Astron.~Obs.~Skalnat{\'e}~Pleso}
\def\pasj{Publ.~Astron.~Soc.~Jpn}
\def\memsai{Mem.~Soc.~Astron.~Ital}
\def\astl{Astron.~Letters}
\def\aipproc{Proc.~AIP~conf.~ser.}
\def\physrep{Physics Reports}
\def\jcap{Journal of Cosmology and Astroparticle Physics}


\begin{thebibliography}{}

\bibitem[\protect\citeauthoryear{{Ahn}, {Iliev}, {Shapiro} \& {Srisawat}}{{Ahn}
  et~al.}{2015}]{Ahn}
{Ahn} K.,  {Iliev} I.~T.,  {Shapiro} P.~R.,    {Srisawat} C.,  2015, \mnras,
  450, 1486

\bibitem[\protect\citeauthoryear{{Angulo}, {Springel}, {White}, {Jenkins},
  {Baugh} \& {Frenk}}{{Angulo} et~al.}{2012}]{Angulo2}
{Angulo} R.~E.,  {Springel} V.,  {White} S.~D.~M.,  {Jenkins} A.,  {Baugh}
  C.~M.,    {Frenk} C.~S.,  2012, \mnras, 426, 2046

\bibitem[\protect\citeauthoryear{{Angulo} \& {White}}{{Angulo} \&
  {White}}{2010}]{angulo_white}
{Angulo} R.~E.,  {White} S.~D.~M.,  2010, \mnras, 405, 143

\bibitem[\protect\citeauthoryear{{Barkana} \& {Loeb}}{{Barkana} \&
  {Loeb}}{2004}]{Barkana04}
{Barkana} R.,  {Loeb} A.,  2004, \apj, 609, 474

\bibitem[\protect\citeauthoryear{{Behroozi}, {Wechsler} \& {Wu}}{{Behroozi}
  et~al.}{2013}]{Behroozi}
{Behroozi} P.~S.,  {Wechsler} R.~H.,    {Wu} H.-Y.,  2013, \apj, 762, 109

\bibitem[\protect\citeauthoryear{{Bhattacharya}, {Heitmann}, {White},
  {Luki{\'c}}, {Wagner} \& {Habib}}{{Bhattacharya} et~al.}{2011}]{bhattacharya}
{Bhattacharya} S.,  {Heitmann} K.,  {White} M.,  {Luki{\'c}} Z.,  {Wagner} C.,
    {Habib} S.,  2011, \apj, 732, 122

\bibitem[\protect\citeauthoryear{{Bond}, {Cole}, {Efstathiou} \&
  {Kaiser}}{{Bond} et~al.}{1991}]{bond}
{Bond} J.~R.,  {Cole} S.,  {Efstathiou} G.,    {Kaiser} N.,  1991, \apj, 379,
  440

\bibitem[\protect\citeauthoryear{{Bower}}{{Bower}}{1991}]{bower}
{Bower} R.~G.,  1991, \mnras, 248, 332

\bibitem[\protect\citeauthoryear{{Boylan-Kolchin}, {Bullock} \&
  {Kaplinghat}}{{Boylan-Kolchin} et~al.}{2012}]{Boylan-Kolchin}
{Boylan-Kolchin} M.,  {Bullock} J.~S.,    {Kaplinghat} M.,  2012, \mnras, 422,
  1203

\bibitem[\protect\citeauthoryear{{Cooray} \& {Sheth}}{{Cooray} \&
  {Sheth}}{2002}]{cooray_sheth}
{Cooray} A.,  {Sheth} R.,  2002, \physrep, 372, 1

\bibitem[\protect\citeauthoryear{{Courtin}, {Rasera}, {Alimi}, {Corasaniti},
  {Boucher} \& {F{\"u}zfa}}{{Courtin} et~al.}{2011}]{Courtin}
{Courtin} J.,  {Rasera} Y.,  {Alimi} J.-M.,  {Corasaniti} P.-S.,  {Boucher} V.,
     {F{\"u}zfa} A.,  2011, \mnras, 410, 1911

\bibitem[\protect\citeauthoryear{{Crocce}, {Fosalba}, {Castander} \&
  {Gazta{\~n}aga}}{{Crocce} et~al.}{2010}]{Crocce}
{Crocce} M.,  {Fosalba} P.,  {Castander} F.~J.,    {Gazta{\~n}aga} E.,  2010,
  \mnras, 403, 1353

\bibitem[\protect\citeauthoryear{{Davis}, {Efstathiou}, {Frenk} \&
  {White}}{{Davis} et~al.}{1985}]{FOF}
{Davis} M.,  {Efstathiou} G.,  {Frenk} C.~S.,    {White} S.~D.~M.,  1985, \apj,
  292, 371

\bibitem[\protect\citeauthoryear{{Del Popolo}}{{Del Popolo}}{2017}]{dp-fit}
{Del Popolo} A.,  2017, Open Astronomy, 26, 26


\bibitem[\protect\citeauthoryear{{Del Popolo} \& {Gambera}}{{Del Popolo} \&
  {Gambera}}{1998}]{DP1}
{Del Popolo} A.,  {Gambera} M.,  1998, \aap, 337, 96

\bibitem[\protect\citeauthoryear{{Del Popolo} \& {Gambera}}{{Del Popolo} \&
  {Gambera}}{1999}]{DP2}
{Del Popolo} A.,  {Gambera} M.,  1999, \aap, 344, 17

\bibitem[\protect\citeauthoryear{{Efstathiou}, {Frenk}, {White} \&
  {Davis}}{{Efstathiou} et~al.}{1988}]{efstat}
{Efstathiou} G.,  {Frenk} C.~S.,  {White} S.~D.~M.,    {Davis} M.,  1988,
  \mnras, 235, 715

\bibitem[\protect\citeauthoryear{{Efstathiou} \& {Rees}}{{Efstathiou} \&
  {Rees}}{1988}]{efstat_rees}
{Efstathiou} G.,  {Rees} M.~J.,  1988, \mnras, 230, 5P

\bibitem[\protect\citeauthoryear{{Gill}, {Knebe} \& {Gibson}}{{Gill}
  et~al.}{2004}]{gill}
{Gill} S.~P.~D.,  {Knebe} A.,    {Gibson} B.~K.,  2004, \mnras, 351, 399

\bibitem[\protect\citeauthoryear{{Gross}}{{Gross}}{1997}]{gross}
{Gross} M.~A.~K.,  1997, PhD thesis, UNIVERSITY OF CALIFORNIA, SANTA CRUZ

\bibitem[\protect\citeauthoryear{{Grossi}, {Dolag}, {Branchini}, {Matarrese} \&
  {Moscardini}}{{Grossi} et~al.}{2007}]{Grossi}
{Grossi} M.,  {Dolag} K.,  {Branchini} E.,  {Matarrese} S.,    {Moscardini} L.,
   2007, \mnras, 382, 1261

\bibitem[\protect\citeauthoryear{{Jenkins}, {Frenk}, {White}, {Colberg},
  {Cole}, {Evrard}, {Couchman} \& {Yoshida}}{{Jenkins} et~al.}{2001}]{jenkins}
{Jenkins} A.,  {Frenk} C.~S.,  {White} S.~D.~M.,  {Colberg} J.~M.,  {Cole} S.,
  {Evrard} A.~E.,  {Couchman} H.~M.~P.,    {Yoshida} N.,  2001, \mnras, 321,
  372

\bibitem[\protect\citeauthoryear{{Kang}, {Norberg} \& {Silk}}{{Kang}
  et~al.}{2007}]{Kang}
{Kang} X.,  {Norberg} P.,    {Silk} J.,  2007, \mnras, 376, 343

\bibitem[\protect\citeauthoryear{{Klypin}, {Trujillo-Gomez} \&
  {Primack}}{{Klypin} et~al.}{2011}]{klypin}
{Klypin} A.~A.,  {Trujillo-Gomez} S.,    {Primack} J.,  2011, \apj, 740, 102

\bibitem[\protect\citeauthoryear{{Knollmann} \& {Knebe}}{{Knollmann} \&
  {Knebe}}{2009}]{knebe}
{Knollmann} S.~R.,  {Knebe} A.,  2009, \apjs, 182, 608

\bibitem[\protect\citeauthoryear{{Lacey} \& {Cole}}{{Lacey} \&
  {Cole}}{1993}]{lacey_cole_93}
{Lacey} C.,  {Cole} S.,  1993, \mnras, 262, 627

\bibitem[\protect\citeauthoryear{{Lacey} \& {Cole}}{{Lacey} \&
  {Cole}}{1994}]{lacey_cole}
{Lacey} C.,  {Cole} S.,  1994, \mnras, 271, 676

\bibitem[\protect\citeauthoryear{{LoVerde}, {Miller}, {Shandera} \&
  {Verde}}{{LoVerde} et~al.}{2008}]{LoVerde}
{LoVerde} M.,  {Miller} A.,  {Shandera} S.,    {Verde} L.,  2008, \jcap, 4, 14

\bibitem[\protect\citeauthoryear{{Mo} \& {White}}{{Mo} \& {White}}{1996}]{Mo}
{Mo} H.~J.,  {White} S.~D.~M.,  1996, \mnras, 282, 347

\bibitem[\protect\citeauthoryear{{Monaco}, {Sefusatti}, {Borgani}, {Crocce},
  {Fosalba}, {Sheth} \& {Theuns}}{{Monaco} et~al.}{2013}]{monaco}
{Monaco} P.,  {Sefusatti} E.,  {Borgani} S.,  {Crocce} M.,  {Fosalba} P.,
  {Sheth} R.~K.,    {Theuns} T.,  2013, \mnras, 433, 2389

\bibitem[\protect\citeauthoryear{{Murray}, {Power} \& {Robotham}}{{Murray}
  et~al.}{2013}]{Murray}
{Murray} S.~G.,  {Power} C.,    {Robotham} A.~S.~G.,  2013, Astronomy and
  Computing, 3, 23

\bibitem[\protect\citeauthoryear{{Press} \& {Schechter}}{{Press} \&
  {Schechter}}{1974}]{PS}
{Press} W.~H.,  {Schechter} P.,  1974, \apj, 187, 425

\bibitem[\protect\citeauthoryear{{Sheth}, {Mo} \& {Tormen}}{{Sheth}
  et~al.}{2001}]{SMT}
{Sheth} R.~K.,  {Mo} H.~J.,    {Tormen} G.,  2001, \mnras, 323, 1

\bibitem[\protect\citeauthoryear{{Sheth} \& {Tormen}}{{Sheth} \&
  {Tormen}}{1999}]{ST}
{Sheth} R.~K.,  {Tormen} G.,  1999, \mnras, 308, 119

\bibitem[\protect\citeauthoryear{{Springel}}{{Springel}}{2005}]{springel}
{Springel} V.,  2005, \mnras, 364, 1105

\bibitem[\protect\citeauthoryear{{Tinker}, {Kravtsov}, {Klypin}, {Abazajian},
  {Warren}, {Yepes}, {Gottl{\"o}ber} \& {Holz}}{{Tinker} et~al.}{2008}]{tinker}
{Tinker} J.,  {Kravtsov} A.~V.,  {Klypin} A.,  {Abazajian} K.,  {Warren} M.,
  {Yepes} G.,  {Gottl{\"o}ber} S.,    {Holz} D.~E.,  2008, \apj, 688, 709

\bibitem[\protect\citeauthoryear{{Viero}, {Moncelsi}, {Mentuch} \& et
  al.}{{Viero} et~al.}{2012}]{viero}
{Viero} M.~P.,  {Moncelsi} L.,  {Mentuch} E.,    et al. 2012, \mnras, 421, 2161

\bibitem[\protect\citeauthoryear{{Vikhlinin}, {Kravtsov}, {Burenin}, {Ebeling},
  {Forman}, {Hornstrup}, {Jones}, {Murray}, {Nagai}, {Quintana} \&
  {Voevodkin}}{{Vikhlinin} et~al.}{2009}]{vikhlinin}
{Vikhlinin} A.,  {Kravtsov} A.~V.,  {Burenin} R.~A.,  {Ebeling} H.,  {Forman}
  W.~R.,  {Hornstrup} A.,  {Jones} C.,  {Murray} S.~S.,  {Nagai} D.,
  {Quintana} H.,    {Voevodkin} A.,  2009, \apj, 692, 1060

\bibitem[\protect\citeauthoryear{{Warren}, {Abazajian}, {Holz} \&
  {Teodoro}}{{Warren} et~al.}{2006}]{warren}
{Warren} M.~S.,  {Abazajian} K.,  {Holz} D.~E.,    {Teodoro} L.,  2006, \apj,
  646, 881

\bibitem[\protect\citeauthoryear{{White}, {Efstathiou} \& {Frenk}}{{White}
  et~al.}{1993}]{white_efstat}
{White} S.~D.~M.,  {Efstathiou} G.,    {Frenk} C.~S.,  1993, \mnras, 262, 1023

\end{thebibliography}

\end{document}